\documentclass[pmlr,twocolumn,10pt]{jmlr}

\usepackage{xurl}
\usepackage{booktabs}
\usepackage{placeins}
\usepackage{siunitx}
\usepackage{float}

\jmlrvolume{}
\jmlryear{}
\jmlrsubmitted{}
\jmlrpublished{}
\jmlrworkshop{}
\makeatletter
\renewcommand*{\@jmlrproceedings}{}
\renewcommand*{\@jmlrabbrvproceedings}{}
\renewcommand*{\@jmlrpages}{}
\makeatother

\title[Paradox of De-id]{\includegraphics[width=0.03\textwidth]{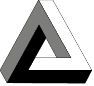} Paradox of De-identification: \linebreak A Critique of HIPAA Safe Harbour in the Age of LLMs}

\author{%
\Name{Lavender Y. Jiang}\Email{lavender.jiang@nyu.edu}
\AND
\Name{Xujin Chris Liu} \Email{xl3942@nyu.edu}
\AND
\Name{Kyunghyun Cho} \Email{kyunghyun.cho@nyu.edu}
\AND
\Name{Eric K. Oermann} \Email{Eric.Oermann@nyulangone.org}\\[1ex]
\addr{\small New York University, New York, NY, USA}
}

\begin{document}

\maketitle

\begin{abstract}
Privacy is a human right that sustains patient-provider trust. Clinical notes capture a patient's private vulnerability and individuality, which are used for care coordination and research. Under HIPAA Safe Harbor, these notes are de-identified to protect patient privacy. However, Safe Harbor was designed for an era of categorical tabular data, focusing on the removal of explicit identifiers while ignoring the latent information found in correlations between identity and quasi-identifiers, which can be captured by modern LLMs. We first formalize these correlations using a causal graph, then validate it empirically through individual re-identification of patients from scrubbed notes. The paradox of de-identification is further shown through a diagnosis ablation: even when all other information is removed, the model can predict the patient's neighborhood based on diagnosis alone. This position paper raises the question of how we can act as a community to uphold patient-provider trust when de-identification is inherently imperfect. We aim to raise awareness and discuss actionable recommendations. 
\end{abstract}

\paragraph*{Data and Code Availability}
From a large, urban, academic hospital, we collected 222,949 identified clinical notes from 170,283 patients (3.34 times more patients than MIMIC-IV, the largest publicly available EHR dataset). Due to the sensitive nature of protected health information and the inherent risks associated with re-identification, neither the raw nor de-identified data can be made publicly available. To support reproducibility, our code is accessible via \url{https://github.com/nyuolab/paradox_of_deidentification}.

\paragraph*{Institutional Review Board (IRB)}
This study was approved by the Institutional Review Board (IRB) at NYU Langone Health (study protocol s21-01189). The methods were carried out in accordance with the IRB's relevant guidelines and regulations.

\begin{figure}[htbp]
\floatconts
  {fig:intro}
  {\caption{Linkage attack of texts is more challenging yet possible with modern language models.}}
  {\includegraphics[width=.9\linewidth]{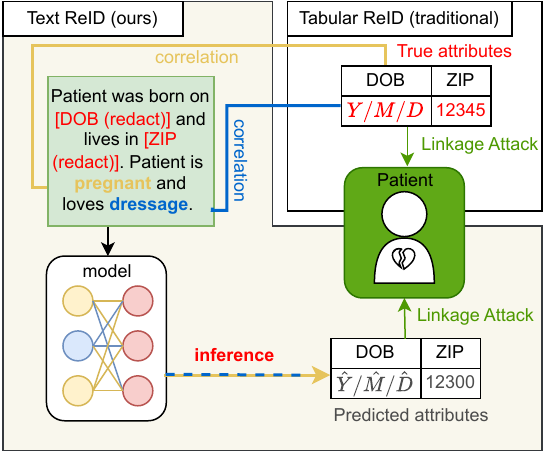}}
\end{figure}

\section{Introduction}
\label{sec:intro}

Privacy is a fundamental pillar of human rights and democratic participation. Although people increasingly sacrifice privacy in exchange for convenience of digital services, studies show that people continue to value privacy and expect protections \citep{Madden2015-mp}.

Healthcare is an exception where people knowingly share the most intimate details of their lives with providers, a disclosure predicated on confidentiality. This is not just a legal requirement but also a clinical necessity, since the quality of care relies on the patient's willingness to be honest \citep{California-Healthcare-Foundation2017-vx, Birkhauer2017-db}. In the United States, this trust is codified by the Health Insurance Portability and Accountability Act (HIPAA).

HIPAA allows for sharing and sale of patient notes after ``de-identification". However, the term increasingly becomes a misnomer since it does not imply the absence of re-identification risk. Under Safe Harbor, ``de-identification" typically involves the removal of 18 categories of Protected Health Information (PHI) \citep{hipaa}.  In practice, this is usually achieved via Named Entity Recognition (NER) and rule-based systems \citep{philter}. 

HIPAA's attempt to balance utility and privacy using scrubbed notes has become an impossible object in the era of LLMs. Much like the Penrose triangle, the redaction appears coherent in isolation but represents a structural paradox when viewed together as a causal graph. Language models can learn nuanced correlations between seemingly non-sensitive information and the patient's identity through backdoor paths that remains after de-identification.

Here we show both principally and empirically that current ``de-identification" practices are insufficient in the presence of LLMs. Through ablation studies, we show that keeping diagnosis alone results in identity leakage. This suggests that ``de-identification" is a paradox: \textbf{how can clinically useful notes be safely shared if the very core medical information enables re-identification}? 

Our goal is to raise awareness and call for culture changes in Health AI community to reflect the capabilities of modern AI. While we do not propose an immediate solution, we posit that the first step toward a remedy is a rigorous redefinition of the problem. 

\section{Related Work}
\paragraph{Linkage Attacks and Regulatory Origins}
The vulnerability of de-identified data to linkage attacks is well-documented. Seminal work by \citet{Sweeney_2000} demonstrated that scrubbed tabular datasets could be linked to individuals using public records, such as the re-identification of the Massachusetts Governor's medical records. These findings directly influenced the design of the HIPAA Safe Harbor provisions, which mandate the removal of specific identifiers to mitigate re-identification risks. However, in the era of Large Language Models (LLMs), the assumption that removing explicit identifiers is sufficient is increasingly challenged. LLMs can exploit latent correlations between non-scrubbed concepts (e.g., stylistic nuances) and the underlying patient identity, weakening the established privacy guarantees.

\paragraph{Critiques of De-Identification}
We are not the first to question the robustness of de-identification. \citet{Ohm_2009} argued that ``data can be either useful or perfectly anonymous but never both,'' describing the ``broken promises of privacy.'' Similarly, the differential privacy community has long argued that traditional de-identification fails to provide rigorous mathematical privacy guarantees \citep{Dwork_Roth_2014}. Despite these academic critiques, the healthcare industry continues to rely heavily on rule-based de-identification tools such as Philter \citep{philter} and treats de-identified clinical notes as a tradable commodity. This suggests a critical gap between theoretical privacy limitations and practical deployment.

\paragraph{Technical Defenses and Legal Frontiers}
Considerable research has focused on improving de-identification techniques, including NER-based approaches to tag explicit identifiers \citep{Neamatullah2008-ug, johnson2016mimic, philter,Yang_Lyu_Lee_Bian_Hogan_Wu_2019,mimic4} and adversarial training and text rewriting \citep{Friedrich_Köhn_Wiedemann_Biemann_2019,Morris_Chiu_Zabih_Rush_2022}. While these technical advancements are important pieces of the puzzles, our work suggests that improving de-identification under the current ``scrub-and-share'' framing may be insufficient. Legal scholarship has also increasingly scrutinized the regulation of highly personal data in the context of LLM interactions \citep{Narayanan2024-mg, Nolte2025-nb,Bommasani2025-et}. We argue that the research community should go beyond minimal legal compliance (specifically the static requirements of Safe Harbor) and adopt a proactive ethical stance that upholds the spirit of patient privacy in the face of increasingly powerful inference technologies.

\section{The paradox from causal graphs}
\label{sec:causal}

\begin{figure*}[htbp]
\centering
\includegraphics[width=.6\linewidth]{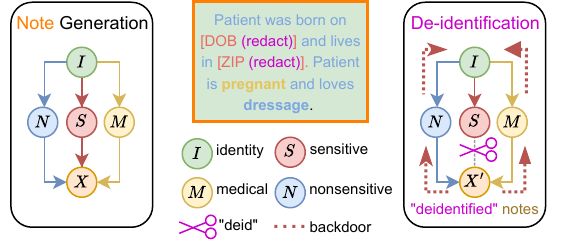}
{\caption{The current de-identification frameworks (purple) leaves two backdoors open for linkage attacks via non-sensitive (blue) and medical (yellow) information.}}\label{fig:causal}
\end{figure*}

By constructing a causal graph using our assumptions on the data-generating process, we can identify the backdoors that are left open for linkage attacks. A causal graph is a type of probabilistic graphical model in which each node is a random variable, and the parents of a node are ``direct causes" of that node. We say $u$ causes $v$, if changes in $u$ result in changes in $v$ when all other variables remain constant. For instance, \autoref{fig:causal} uses $I$ to represent the identity of a patient (green) and $M$ to represent the patient's medical information (yellow). For illustration, let's assume $M$ represents pregnancy status. If we change the patient's biological sex from female to male, then $M$ must be ``not pregnant". Therefore, we draw a directed edge from $I$ to $M$ that encodes a causation relationship.

\paragraph{Note Generation} 
\autoref{fig:causal} (left) illustrates the generating process of real-world clinical notes based on our assumptions. The green node $I$ represents the patient's identity (e.g., address and occupation). Associated with this identity are sensitive information $S$ (red node; e.g., DOB and ZIP code), medical information $M$ (yellow node; e.g., diagnosis), and non-sensitive information $N$ (blue node; e.g., a hobby such as breakdancing). The clinical note $X$ (orange node) is derived from these three variables ($S, M, N$). We assume that only the clinical note $X$ is observed and shared, while the other latent variables remain unobserved. 

\paragraph{De-identification}
\autoref{fig:causal} (right) illustrates how the current de-identification paradigm leaves two backdoors open for linkage attacks. This paradigm produces ``de-identified'' notes $X'$ by detecting and removing HIPAA-protected attributes, effectively severing the red edge from $S$ to $X$. However, the correlation between $X'$ and $I$ persists, mediated by backdoor paths (represented by dotted red arrows). First, the blue path ($X'\leftarrow N\leftarrow I \rightarrow S$) shows the correlation via non-sensitive attributes $N$. Second, the yellow path ($X'\leftarrow M\leftarrow I \rightarrow S$) shows the correlation via medical information $M$. 

\paragraph{Examples}
To make this concrete, let's examine both a synthetic and a real-world example. Consider the clinical note snippet in the center of \autoref{fig:causal}: ``Patient was born on [DOB] and lives at [ZIP code]. Patient is pregnant and loves dressage.'' Although the protected attributes (DOB and ZIP code) are redacted, we can still infer that the patient is an adult female based on the pregnancy, and resides in an affluent neighborhood given the hobby of dressage. Regarding real-world data, consider MIMIC-III and MIMIC-IV, which were de-identified by masking out HIPAA-protected attributes \citep{johnson2016mimic, mimic4}. These datasets contain 7-12 years of electronic health records (EHR) from the ICU (or ICU and ED) of Beth Israel Hospital, covering 38,597 - 50,920 adult patients. De-identification in these datasets employed regular expression filters, language model-based detectors, or a combination of both to detect and replace sensitive attributes such as name and address. 

\begin{table*}[htbp!]
\centering
\begin{tabular}{l p{50mm} S[table-format=2.2]}
\toprule
\bfseries Data & \bfseries Path & {\bfseries AUC} \\ 
\midrule
Random Guess & None & 50.00 \\ 
\addlinespace[0.3em]
Diagnosis Only & $X'\rightarrow M\leftarrow I \rightarrow S$ only & 58.57 \\ 
\addlinespace[0.3em]
De-identified Note & $X'\rightarrow M\leftarrow I \rightarrow S$ and \newline $X'\rightarrow N\leftarrow I \rightarrow S$ & 78.35 \\ 
\addlinespace[0.3em]
Identified Note & All open paths (via $M,N,S$) & 82.78 \\ 
\bottomrule
\end{tabular}
\caption{Borough prediction AUC shows that diagnosis alone leaks identity, with non-sensitive information contributing most to re-identification risk.}
\label{tab:condition-only}
\end{table*}

\paragraph{The Paradox} 
Our causal graph reveals that ``de-identification'' is fundamentally a misnomer. Current practices define success by severing the direct link ($S\rightarrow X$) via identifier removal, while systematically ignoring backdoor paths persisting through non-sensitive ($N$) and medical ($M$) attributes. This creates an inescapable paradox: even under perfect Safe Harbor compliance, ``de-identified'' notes remain statistically tethered to identity through the very correlations that confirm their clinical utility. The conflict is structural instead of technical. While \citet{Scaiano2016-hp} warned of quasi-identifiers, we argue the tension is intrinsic: complete privacy demands severing all pathways to identity, but clinical utility requires preserving the medical content that leaks it. \autoref{tab:condition-only} exposes this paradox empirically. A model predicting patient neighborhood achieves above-random accuracy from diagnosis alone (AUC 58.57\%), rising to 78.35\% with de-identified notes, which is startlingly close to the 82.78\% achieved with fully identified notes. This narrow 4.43 point gap reveals a troubling truth: the vast majority of re-identification risk stems not from Protected Health Information, but from the non-sensitive and medical content we deem safe to share. Appendix~\ref{apd:maps} visualizes this geographic variation, showing that health outcomes and demographics cluster by neighborhood. See section \ref{sec:exp_setup} for experimental details.

\section{Empirical Evidence from Re-identification}
\label{sec:exp}

The backdoor paths identified in \autoref{sec:causal} suggest that patients can be re-identified from standard de-identified notes with accuracy significantly exceeding random chance. To validate this, we conduct a two-stage attack: first inferring sensitive attributes from redacted notes, and then using these predictions to link \textit{individuals}. This approach goes beyond typical privacy studies by demonstrating actual individual linkage rather than just attribute inference. We benchmark performance against random baselines: for attribute prediction, the baseline is uniform probability. For re-identification, the baseline consists of guessing the majority class of each attribute and drawing uniformly from the matched group.

\begin{figure*}[t] 
\centering
\includegraphics[width=\textwidth]{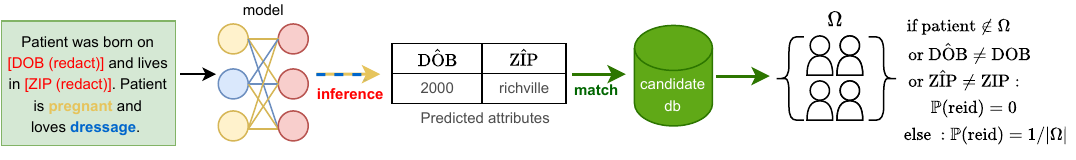}
\caption{\textbf{Re-identification process overview.} As a simplified example, a single de-identified notes is passed through a LM for DOB and ZIP prediction. The predictions are used to match patients in a candidate database, resulting in a candidate set $\Omega$. In this case, the re-identification risk is zero if either of the prediction is wrong. Otherwise, the risk is $1/|\Omega|$ based on random draw from the correctly identified group.}
\label{fig:reid}
\end{figure*}

\autoref{fig:reid} illustrates the re-identification attack process. First, an adversary trains a model to learn correlations between sensitive attributes and the remaining content (medical and non-sensitive) in de-identified notes. Second, the adversary applies this model to unseen notes to predict the redacted sensitive attributes (e.g., DOB and zip code). Finally, using the predicted attributes, the adversary queries an external database to identify individuals with matching attributes and randomly selects one person from the matching group. If the selected person is the original patient, the attack succeeds; otherwise, it fails. 

For simplicity, we assume the adversary has access to the complete patient population in the external database. In practice, the adversary may only have access to a subset of the population, which would make the attack more difficult. However, even with this assumption, we can still demonstrate that de-identified notes are not as private as they seem.

\subsection{Experiment Setup}\label{sec:exp_setup}

\paragraph{Data.} We collected 222,949 identified clinical notes from 170,283 patients at NYU Langone, a large, urban, academic hospital. This cohort is 3.34 times larger than MIMIC-IV, the largest publicly available EHR dataset. For each patient, we curated six demographic attributes: biological sex, note year, note month, borough, zip code median income, and insurance type (see \autoref{tab:attributes}). These attributes were selected to approximate the unique identifier trio \citep{Sweeney_2000} consisting of biological sex, birth date, and zip code, which can collectively isolate over 87\% of the U.S. population. In our work, birth date is proxied by note year and month (combined with age mentions), and zip code is inferred from borough, area income, and insurance type (see Appendix~\ref{apd:maps} for geographic distributions of these attributes). Clinical notes were de-identified using \textsc{UCSF philter} \citep{philter}. We partitioned the dataset by patient into training (80\%), validation (10\%), and test (10\%) splits, ensuring all notes from a given patient remain in the same split.

\begin{table*}[htp]
\centering
\begin{tabular}{lcl}
\toprule
Attribute           & \# Classes & Potential Values                                                     \\ 
\midrule
Biological Sex                 & 2               & Female, Male                                              \\ 
Neighborhood        & 6               & Manhattan, Brooklyn, Bronx, Queens, Staten Island, Others\\ 
Note Year           & 10              & 2012--2021                                                \\ 
Note Month          & 12              & January--December                                         \\ 
Area Income         & 2               & Poor (below median), Rich (above median)                  \\ 
Insurance Type      & 2               & Public (Medicare/Medicaid), Private                       \\
\bottomrule
\end{tabular}
\caption{Demographic attributes predicted from de-identified clinical notes. These attributes approximate the unique identifier trio from \citet{Sweeney_2000}.}
\label{tab:attributes}
\end{table*}

\paragraph{Finetuning} We fine-tuned BERT-base-uncased (110M parameters) \citep{Devlin_Chang_Lee_Toutanova_2018}, pretrained only on general text to prevent clinical information leakage, to predict each attribute separately. See Appendix \ref{apd:ft} for details.

\paragraph{Prediction Evaluation.} We evaluate the generalization capability of our fine-tuned models on the test set using two complementary metrics: Accuracy and weighted Area Under the Receiver Operating Characteristic Curve (ROC-AUC). While Accuracy measures the absolute hit rate, ROC-AUC provides a robust estimate of discriminative power, particularly valuable for attributes with class imbalance.

\paragraph{Re-identification Strategy.} We simulate an adversarial linkage attack where fine-tuned models recover redacted demographic attributes from de-identified notes to pinpoint individuals in a patient database. We employ a top-$k$ matching strategy: for the $i$-th attribute, we select the top $k_i$ predicted classes and filter the database for patients matching any of these values. To rigorously analyze the trade-off between prediction accuracy and the specificity of the matched group, we exhaustively sweep $k_i \in \{1, \dots, c_i\}$ for attributes with $c_i$ classes.

\paragraph{Re-identification Evaluation.} We assess risk using three metrics: 
\begin{enumerate}
\item \textbf{Group Re-identification Success Rate} ($\mathbb{P}_G=\Pi_{a\in A} \mathbb{P}_{a}$): the frequency with which the true patient is captured within the predicted candidate set $\Omega$. This requires correct top-$k$ predictions for all attributes $a \in A$ such as neighborhood.
\item \textbf{Individual Re-identification from Group} ($\mathbb{P}_{I\mid G}$), the probability of identifying the true patient from a correctly identified group
\item \textbf{Probability of Unique Re-identification} ($\mathbb{P}_{\mathrm{reid}}=\mathbb{P}_{I\mid G}\times\mathbb{P}_{G}$), the overall likelihood that an adversary uniquely identify a patient from de-identified notes.
\end{enumerate}

Note that $\mathbb{P}_{\mathrm{reid}}$ is a conservative lower bound because we model $P_{I\mid G}$ as a uniform $1/|\Omega|$ chance, where $|\Omega|$ is the size of the candidate pool. This however likely underestimates real-world risks, because adversaries often leverage auxiliary knowledge such as social media to prune the candidate pool and achieve higher individual matching accuracy.

\subsection{Experimental Results}
\begin{figure*}[htp]
    \centering
    \includegraphics[width=.9\linewidth]{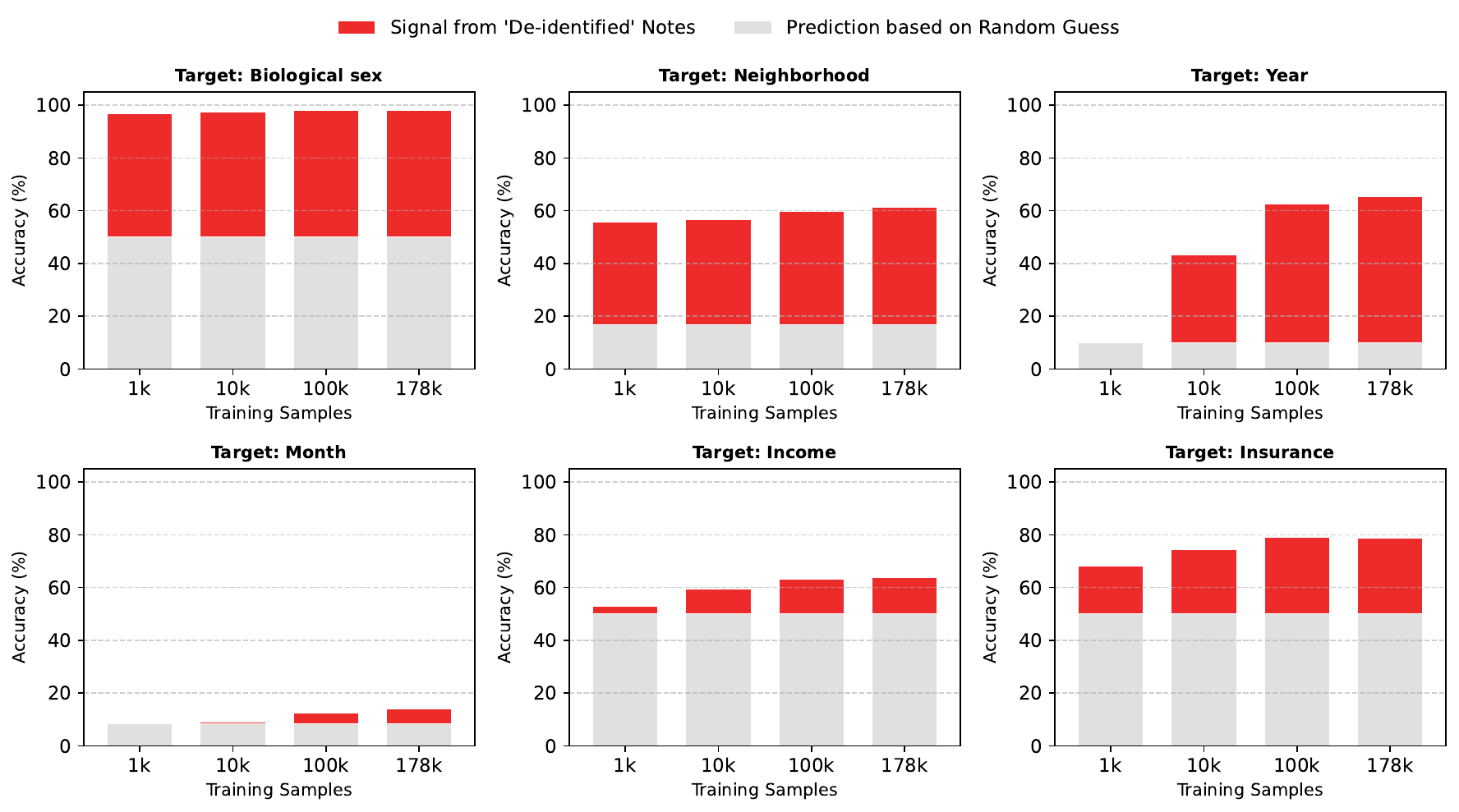}
    \caption{\textbf{Attribute prediction exceeds random chance using as few as 1k examples.} LM's accuracy (red bars) is better than random guess (grey bars) and improve with more data. AUCs shows a similar above-random pattern whose gap increases with more data (Appendix \ref{apd:auc_bar}).}
    \label{fig:bar_acc}
\end{figure*}

\paragraph{Attribute prediction exceeds random chance using as few as one thousand examples.}\label{subsec:finetuned_perf}
As illustrated in \autoref{fig:bar_acc}, de-identified clinical notes remain vulnerable to attribute prediction. Across all six attributes and all data regimes (1k to 177k examples), the language model (red) consistently outperform random baselines (grey). These results empirically supports that de-identification process retains exploitable signals in the two backdoor paths. 

The privacy risk is immediate: models achieve above-random performance with as few as 1,000 training examples. While biological sex is the most exposed attribute (recovered with $>99.7\%$ accuracy), even the subtlest signals (month of note) are predicted with better-than-random accuracy.

\begin{figure*}[htp]
\centering
\centering
\includegraphics[width=\linewidth]{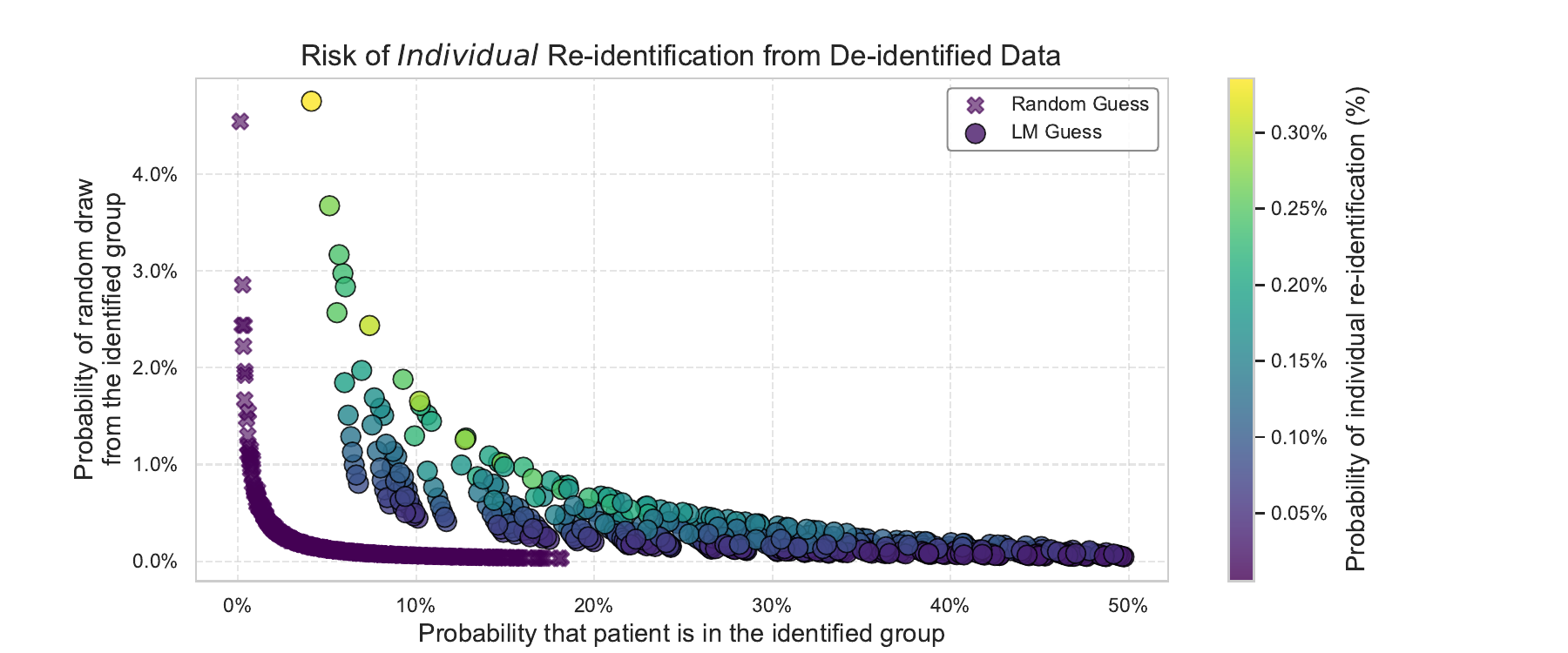}
\caption{\textbf{Individual re-identification risk significantly exceeds random guess.} The scatter plot shows the relationship between group re-identification probability($\mathbb{P}_G$, $x$ axis) and the the probability of identifying the correct individual within that group ($\mathbb{P}_{I\mid G}\approx 1 / |\Omega|$, $y$ axis). More yellow points toward the upper right represents higher overall re-identification risks. LM predictions (circles) consistently outperform majority class guesses (crosses).}
\label{fig:acc_group_size_plot}
\end{figure*}

\paragraph{\textit{Individual} re-identification significantly exceeds random baseline.} We evaluate the risk of individual re-identification  ($\mathbb{P}_\mathrm{reid} = \mathbb{P}_{I\mid G} \times \mathbb{P}_G$) as the product of the probability of correct group identification ($\mathbb{P}_G$) and the probability of selecting the true patient from that identified group ($\mathbb{P}_{I\mid G} \approx 1 / |\Omega|$). \autoref{fig:acc_group_size_plot} visualizes this relationship, comparing language model inferences (circles) against guessing majority-class (crosses). 

The language model predictions shift clearly toward the top-right of the plot. The top direction shows that for any given group success rate ($\mathbb{P}_G$), LM identifies smaller group (higher $\mathbb{P}_{I\mid G}$). The right direction shows that for any fixed identified group size ($\mathbb{P}_{I\mid G}$), LM has a higher group success rate ($P_G$). In fact, the max re-identification risk for the language model ($0.34\%$) is approximately 37 times higher than the highest majority-class baseline ($0.0091\%$).

\paragraph{Interpretation of the risks.} 
While a $0.34\%$ re-identification risk may appear numerically small, the determination of ``acceptable" risk is a complex policy question that statistics alone cannot resolve. For patients with rare medical histories or marginalized identities, even a low probability of exposure constitutes a significant breach of safety and trust. Furthermore, this risk should be viewed at scale: when applied to a corpus the size of MIMIC-IV, a $0.34\%$ risk implies the potential re-identification of roughly 170 patients. The question then becomes how many leaks are acceptable, particularly when data subjects are not aware of the possibility of re-identification via secondary model inference.

\section{A Call to Action}
Our findings challenge the core premise of the HIPAA Safe Harbor standard, which operates on a binary definition of privacy: data is either ``identified'' or ``de-identified.'' HIPAA assumes that removing a static list of tokens renders data ``safe", effectively decoupling the clinical narrative from the patient's identity. However, our causal graph analysis and empirical results suggests that this decoupling is a mirage. Clinical notes are inherently entangled with identity. A patient's medical diagnosis and non-redacted narratives are direct products of their unique life trajectory, creating high-dimensional signature that can be mapped back to the individual.

By treating de-identification as a deterministic process of redaction, the current framework ignores the correlation remaining in the residual text. In the era of LLMs, which excel at synthesizing subtle and heterogeneous patterns, the distinction between ``identifiers'' and ``content'' collapses. A specific diagnosis paired with a niche hobby is no longer just medical data but also a smeared fingerprint.

This reveals a sociotechnical disconnect: policy accepts a method (scrubbing 18 identifiers) as a proxy for a goal (anonymity), but advancing technology continuously degrades that proxy's validity. If we accept a 0.34\% re-identification risk as ``safe'', we are implicitly approving the exposure of over 800,000 individuals, a figure derived from applying our empirically observed rate (measured on a diverse urban hospital population) to the 278 million US urban residents. For vulnerable populations, this is not a statistic but a breach of trust. The commodification of ``de-identified'' data thus rests on a precarious ethical foundation, trading on a promise of privacy that modern LM is uniquely equipped to break.

\subsection{The Urgency}

The acceleration of the AI arms race has transformed high-quality domain-specific text into one of the world's most sought-after commodities. LLMs are data-hungry, requiring trillions of tokens to achieve emergent capabilities \citep{Groeneveld2024-py,Dubey2024-qg}. As the low-hanging fruit of public web data is exhausted, protected and high-quality corpora such as clinical notes become the next frontier. 

Healthcare already accounts for about 30\% of the global data volume \citep{healthcare_data}, yet it remains relatively untouched due to privacy constraints. However, this is ending as major AI companies such as OpenAI and Anthropic pivot towards clinical integration \citep{OpenAI2026-lv, OpenAI2026-ur, Anthropic2026-kr}. The current legal framework relies on Safe Harbor to justify healthcare data transfer, but our results demonstrate that its privacy guarantee is insufficient, especially in the face of modern LLMs. 

The persistence of Safe Harbor despite known limitations is not an oversight but a feature of a system optimized for data liquidity rather than patient protection \citep{Ohm_2009}. De-identified clinical notes represent a multi-billion dollar market \citep{gvr-pw}, creating structural disincentives for healthcare institutions to adopt privacy-preserving alternatives that might reduce data utility or require costly infrastructure investments. There is an urgency to carefully investigate, understand and address this disincentive.

\subsection{Policy Recommendations} 
We propose a shift from binary privacy definitions to a risk-aware governance framework:

\paragraph{Tiered Access and Accountability:} We call for a regulatory framework where access controls are proportional to re-identification risk. High-risk data necessitates stringent vetting of researchers and regular, renewable access audits by trusted third parties. Furthermore, shared datasets should implement digital watermarking to ensure provenance and traceability, discouraging negligence.

\paragraph{Transparency and Patient Rights:} Patients should be informed that ``de-identification" is a probabilistic risk reduction, not a guarantee of anonymity. We advocate for the right to know when and how one's data is being utilized, shifting the paradigm from implicit waivers to active engagement.
 
\paragraph{Quantifiable Utility Checks:} Data release should be governed by rigorous pre-sharing evaluations. We recommend adopting utility quantification metrics such as SecureKL \citep{Fuentes2025-lh} to ensure that privacy risks are justified by tangible scientific value. Data that fails to meet utility thresholds should not be exposed to privacy risks.

\subsection{Research Recommendations} 

Safeguarding sensitive clinical data requires moving beyond a purely technocentric view. We call for interdisciplinary research that bridges engineering with legal and social governance.

\paragraph{Rejecting Technical Hubris:} It is time to abandon the assumption that de-identification is a problem solvable through engineering alone. As our causal analysis demonstrates, perfect de-identification of high-dimensional clinical text is a technically impossible goal. Research should pivot towards co-designing systems where technical safeguards are reinforced by legal liability and social contracts, rather than expecting software to replace the rule of law.

\paragraph{Advancing De-identification Standards:} While technical solutions alone are insufficient, they remain a critical line of defense. Future works should move beyond heuristics to more principled approaches. Promising directions include differentially private synthetic data generation \citep{Near_2021,Li_2022,Lin_Gopi_Kulkarni_Nori_Yekhanin_2023}.

\paragraph{Cultivating Data Transparency:} We advocate for the adoption of ``data nutrition labels" \citep{Sun2019-gj} to transparently communicate the privacy limitations of datasets. We should further recognize that privacy risks propagate to downstream models. Since language models can memorize training data \citep{Carlini_Ippolito_Jagielski_Lee_Tramer_Zhang_2022}, models finetuned on ``de-identified" or synthetic data should be subject to the same privacy considerations as the underlying data.

\section{Conclusion}
Medical privacy is foundational to patient-provider trust, quality of care, and individual dignity. Currently, this protection relies on HIPAA Safe Harbor, a standard we argue is increasingly fragile in the era of LLMs. By reducing privacy to the removal of 18 explicit identifiers, current regulations fail to account for the high-dimensional correlations that permeate clinical narratives. We frame de-identification as a paradox: much like a Penrose triangle, the redaction process appears sound locally but proves structurally impossible when viewed as a whole. Through causal analysis and empirical validation (including neighborhood prediction from diagnosis and an end-to-end re-identification attack on $\sim$17k patients), we demonstrate that seemingly de-identified notes retain nontrivial signals for individual re-identification. As data-hungry AI systems expand into healthcare, the urgency of addressing this vulnerability cannot be overstated. We offer concrete recommendations for policymakers and researchers to move beyond ``scrub-and-share'' models. While no immediate technical panacea exists, acknowledging this paradox is the first step toward a more robust framework for data governance and patient protection.

\section{Limitations}

Our re-identification attack assumes access to labeled training data to learn the correlations between redacted notes and sensitive attributes. While our study utilized internal hospital records, real-world adversaries can leverage vast repositories of public and illicitly obtained data. For instance, voter registration lists publicly disclose names, addresses (including zip codes), birth dates, biological sex, and political affiliations. Furthermore, sensitive personal data is increasingly exposed through frequent data breaches \citep{Landi2020-ww, Bruce2024-aj, Office-for-Civil-Rights-OCR-2024-cp}, providing adversaries with rich auxiliary datasets for linkage attacks.

This study purposefully establishes a conservative baseline by simplifying the experimental setting. In practice, the re-identification risk is likely higher, motivated adversaries could employ more sophisticated techniques for finding the true patient in an identified group. For instance, an adversary could deploy autonomous agent built on trillion-parameter backbones, capable of cross-referencing clinical notes with live web search, social media footprints, and other digital traces to achieve high-precision re-identification. 

Finally, our empirical evaluation is restricted to a single healthcare system. Although the study is limited to data from a single health system, this system comprises three distinct locations with highly diverse patient demographic. While evaluating multi-institutional datasets could further strengthen our empirical results, the marginal utility of such data access is outweighed by the ethical and privacy risks associated with accessing additional raw identified records. Furthermore, our theoretical framework (illustrated by \autoref{fig:causal}) shows that the persistence of backdoor paths is a structural property of the redaction process itself. Given that the feasibility of linkage attack is well-documented in the literature, these structural vulnerabilities suggest that our findings are broadly applicable to other clinical environments utilizing similar HIPAA Safe Harbor protocols.

\acks{L.Y.J. is supported by Apple AIML PhD fellowship. L.Y.J. and K.C. are supported by NSF Award 1922658. K.C., E.K.O. and L.Y.J. are supported by Institute for Information \& communications Technology Promotion (IITP) grant funded by the Korea government (MSIT) (No. RS-2019-II190075 Artificial Intelligence Graduate School Program (KAIST); No. RS-2024-00509279, Global AI Frontier Lab). E.K.O. is supported by the National Cancer Institute's Early Surgeon Scientist Program (3P30CA016087-41S1) and the W.M. Keck Foundation. We would like to thank Mimee Xu, Lucas Rosenblatt, Daniel Alber, Gavin Zihao Yang, Karl Lee Sangwon, Zachary Horvitz, 
Hilal Asi, Julia Stoyanovich, Saadia Gabriel, A. Feder Cooper, Hima Lakkaraju, Angelica Chen, Stephanie Milani, Irene Chen, Falaah Arif Khan, Alice Oh for valuable discussion. }

Gemini-3-pro and GPT-5.2 were used for revising manuscript and improving figure. Claude Opus 4.6 were used for resolving outdated dependencies for open-source code and improving documentation.

\paragraph{Author's contribution}
K.C. and E.K.O. supervised the project.  L.Y.J., K.C. and E.K.O. conceptualized the project. L.Y.J collected data and engineered the software for training and evaluation. L.Y.J., X.C.L and K.C. debug the software. L.Y.J., K.C., X.C.L., E.K.O. created figures. L.Y.J., X.C.L. and K.C. wrote the initial draft. All authors edited and revised the manuscript.

\clearpage
\bibliography{chil-sample}

\clearpage

\appendix

\begin{figure*}[htp]
    \centering
    \includegraphics[width=\linewidth]{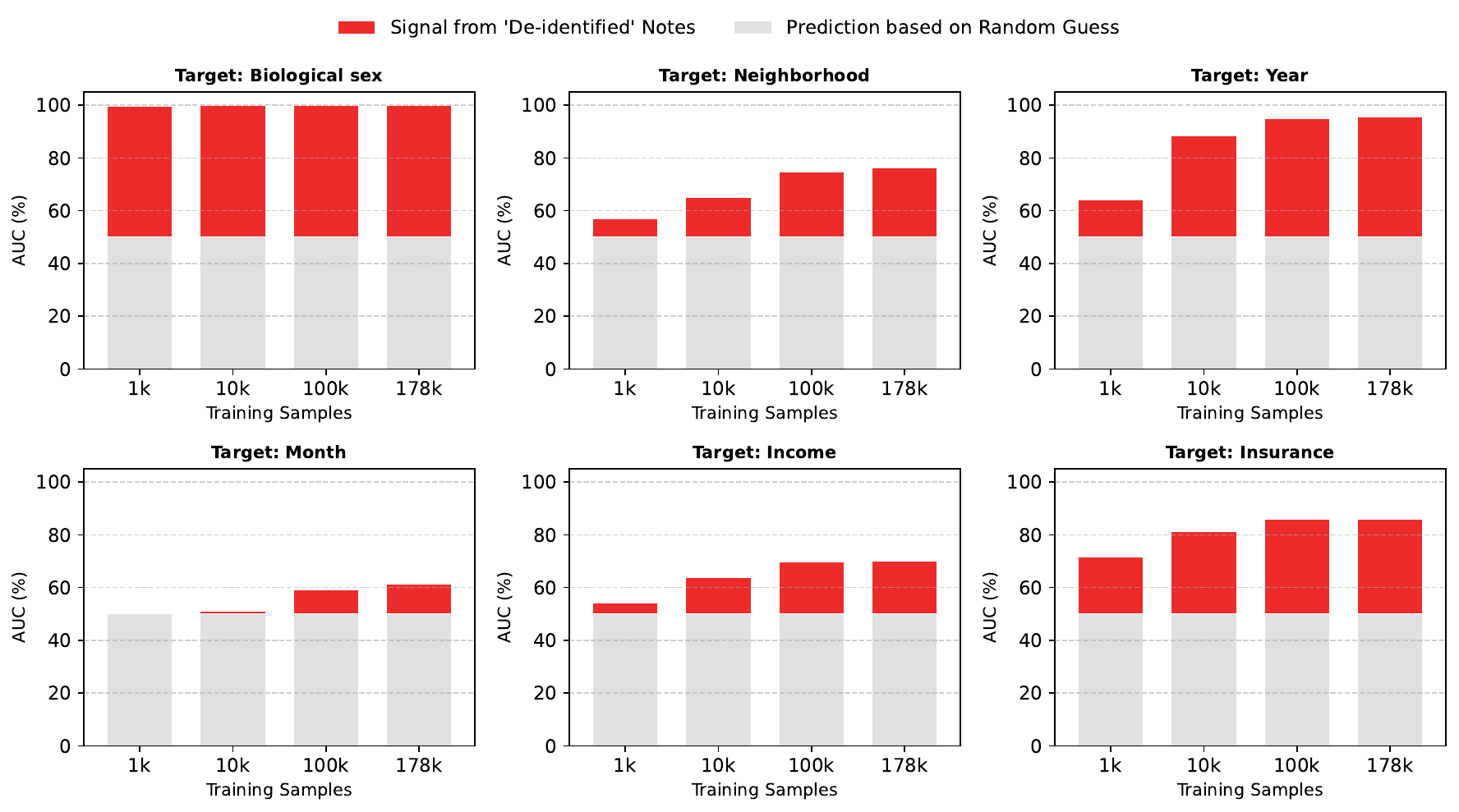}
    \caption{Attribute predictor's AUC (red bars) is better than random guess (blue bars).}
    \label{fig:bar_auc}
\end{figure*}

\FloatBarrier

\section{Above-random AUCs}\label{apd:auc_bar}

\autoref{fig:bar_auc} shows that similar to accuracy, the AUC of each attribute predictor is above random guess and improves with more training data.

\section{Training Details}\label{apd:ft}

\paragraph{Finetuning.} To prevent information leakage from clinical pre-training, we initialized our approach with \textsc{bert-base-uncased} (110M parameters), which is pre-trained solely on general domain text \citep{Devlin_Chang_Lee_Toutanova_2018}. We fine-tuned a separate model for each attribute using eight NVIDIA A100 GPUs (40GB) or H100 GPUs (80GB) for up to 10 epochs, with early stopping and a random seed of 0. We used the AdamW optimizer \citep{Loshchilov_Hutter_2017} with a learning rate of \num{2e-5}, no weight decay, an effective batch size of 256, and a linear decay schedule without warmup. Checkpoints were evaluated every half-epoch, and the model achieving the highest weighted validation ROC-AUC was selected for inference. Code is available on Anonymous Github.

\section{Geographic Distribution of Health Outcomes}\label{apd:maps}

\begin{figure*}[htp]
    \centering
    \includegraphics[width=0.48\linewidth]{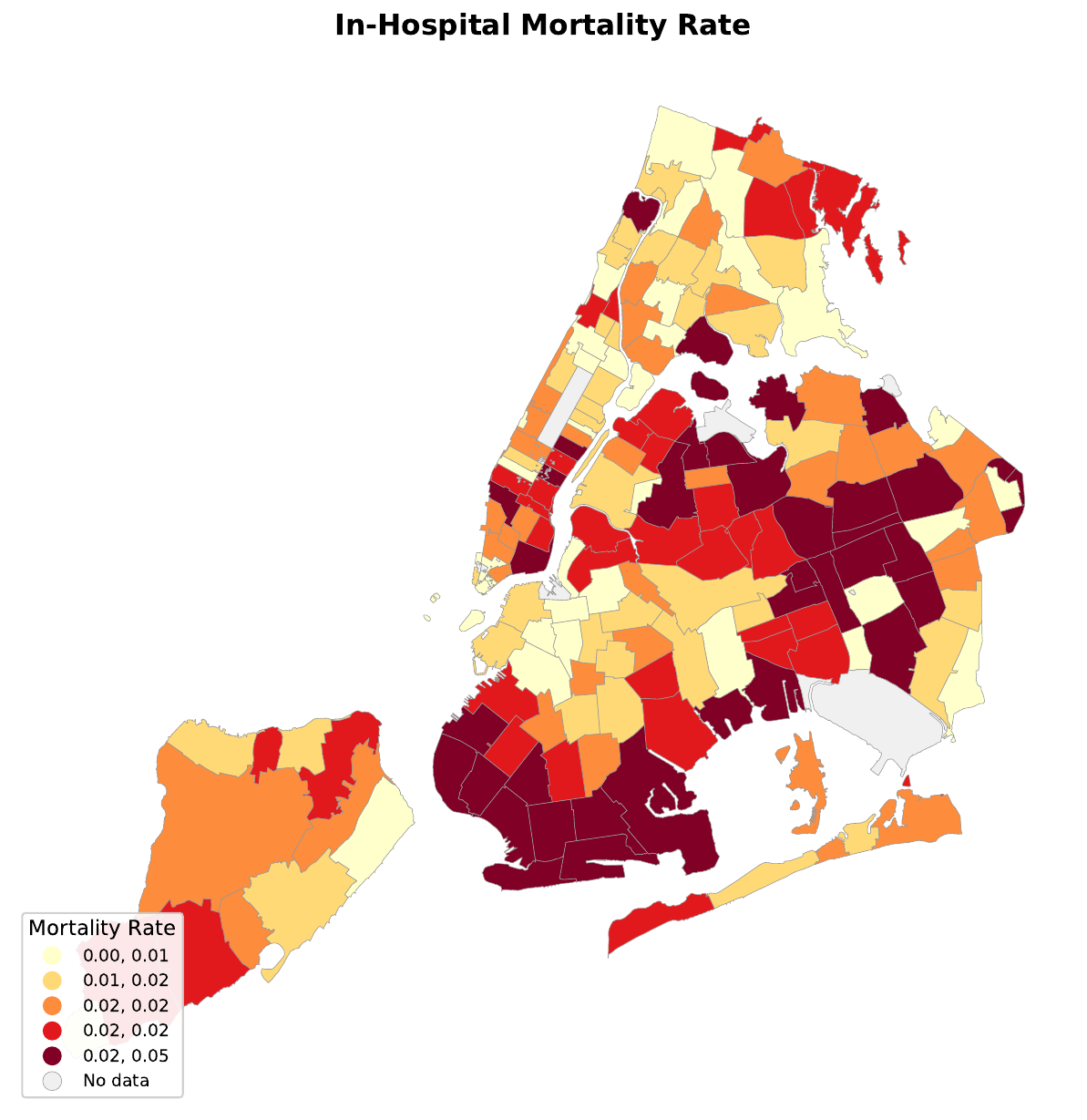}
    \hfill
    \includegraphics[width=0.48\linewidth]{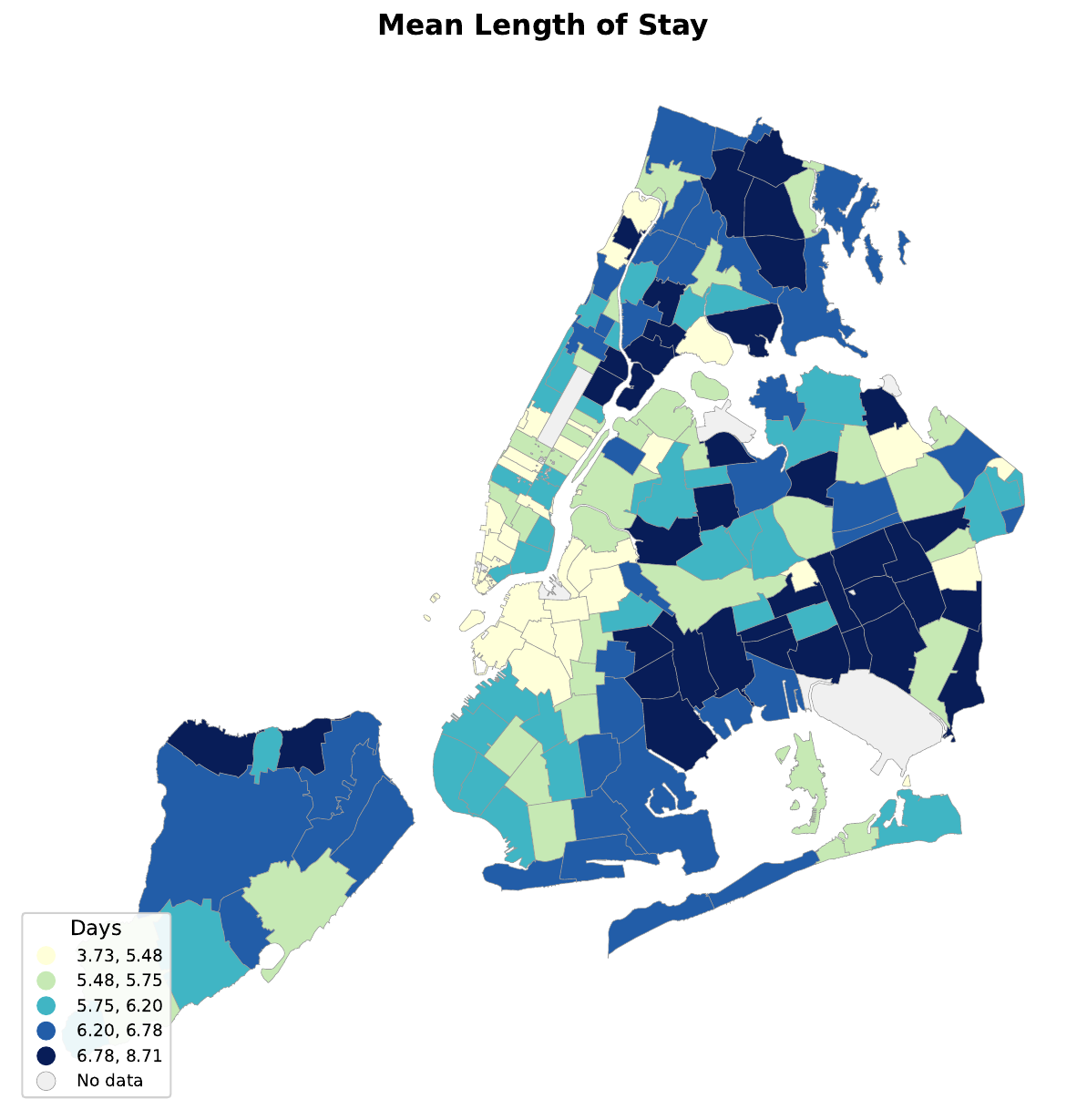}
    \caption{NYC zip code heatmaps of in-hospital mortality rate (left) and mean length of stay (right).}
    \label{fig:nyc_maps_outcomes}
\end{figure*}

\begin{figure*}[htp]
    \centering
    \includegraphics[width=0.48\linewidth]{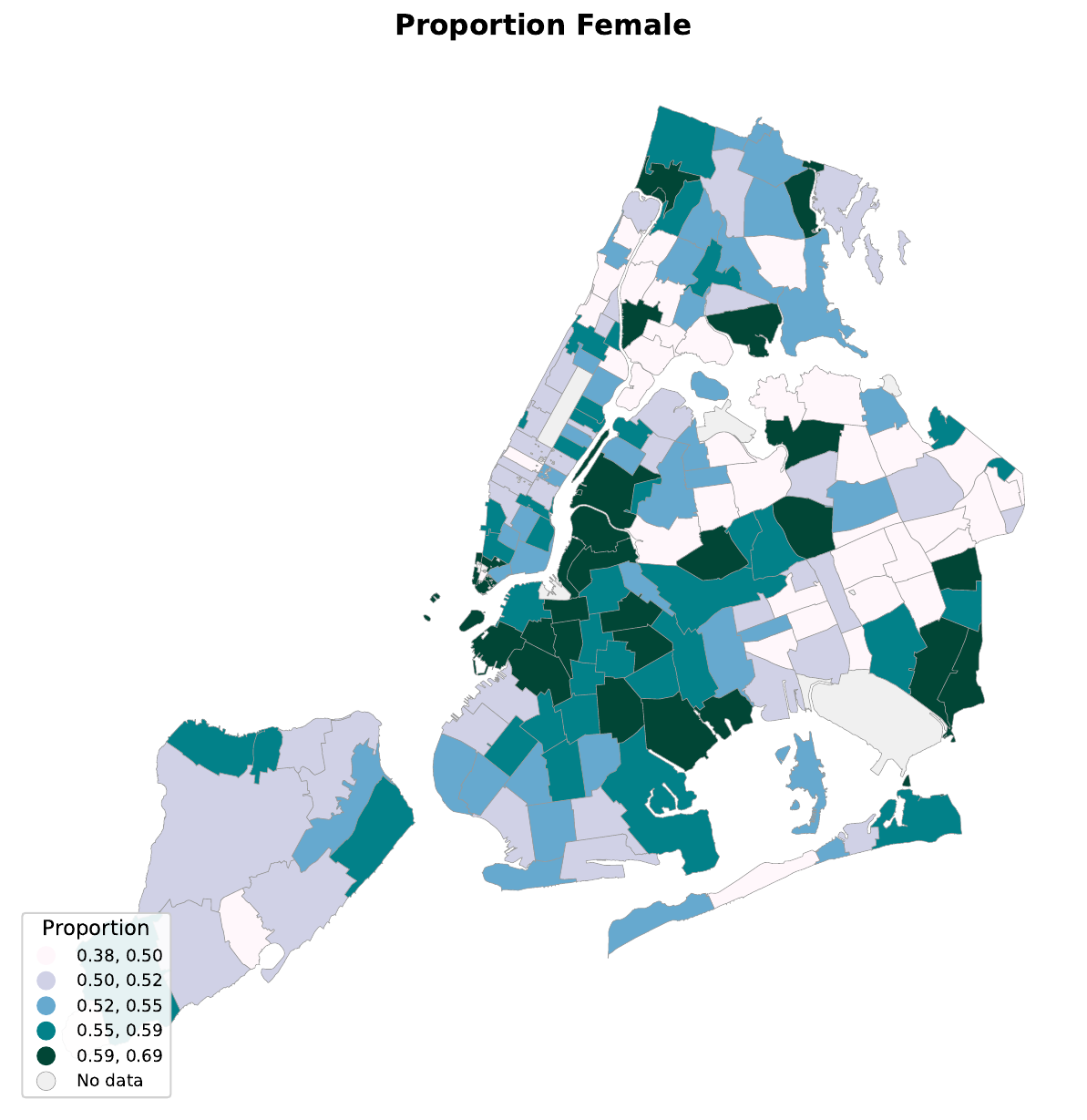}
    \hfill
    \includegraphics[width=0.48\linewidth]{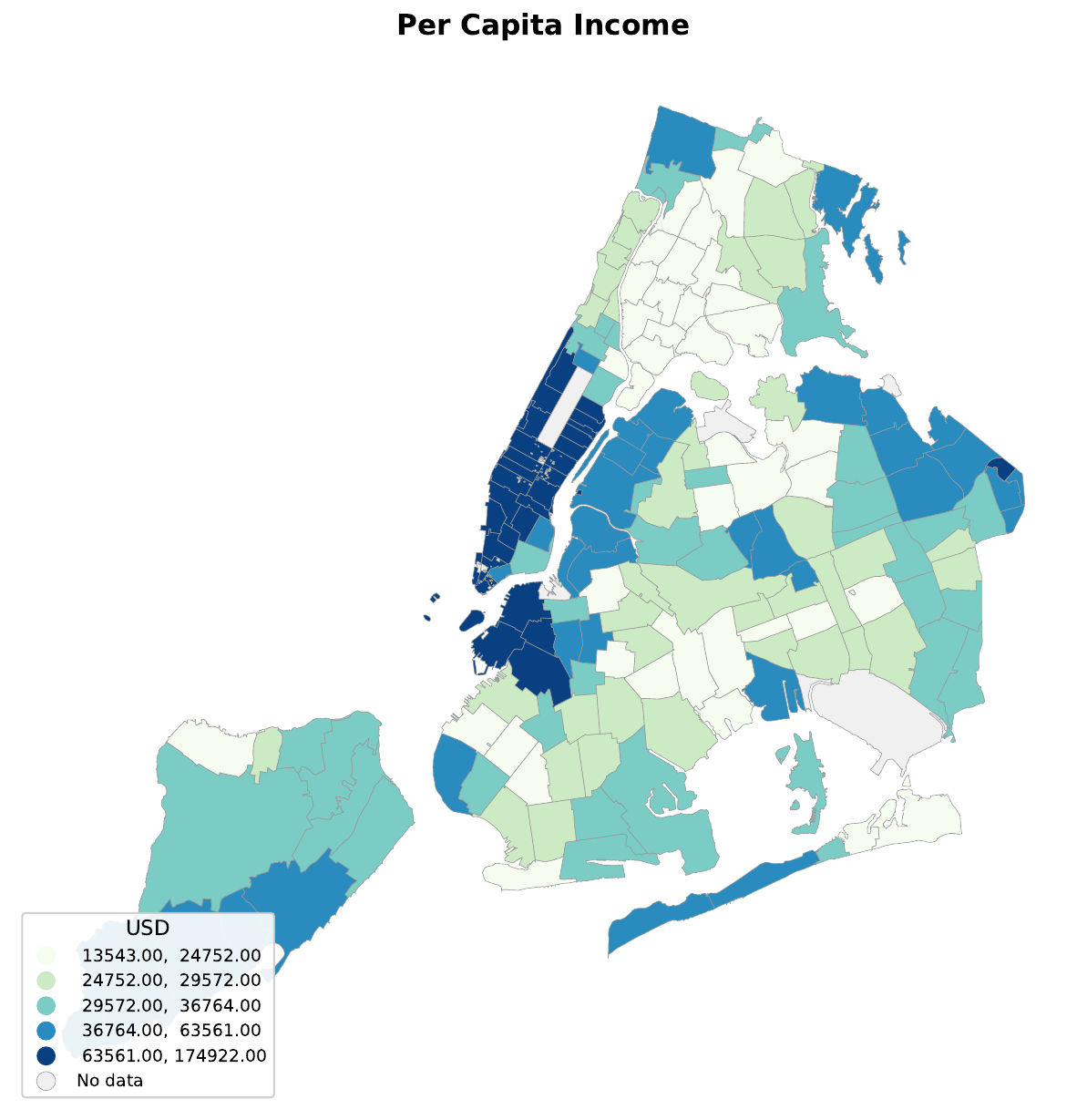}
    \caption{NYC zip code heatmaps of proportion female (left) and per capita income (right).}
    \label{fig:nyc_maps_demo1}
\end{figure*}

\begin{figure*}[htp]
    \centering
    \includegraphics[width=0.48\linewidth]{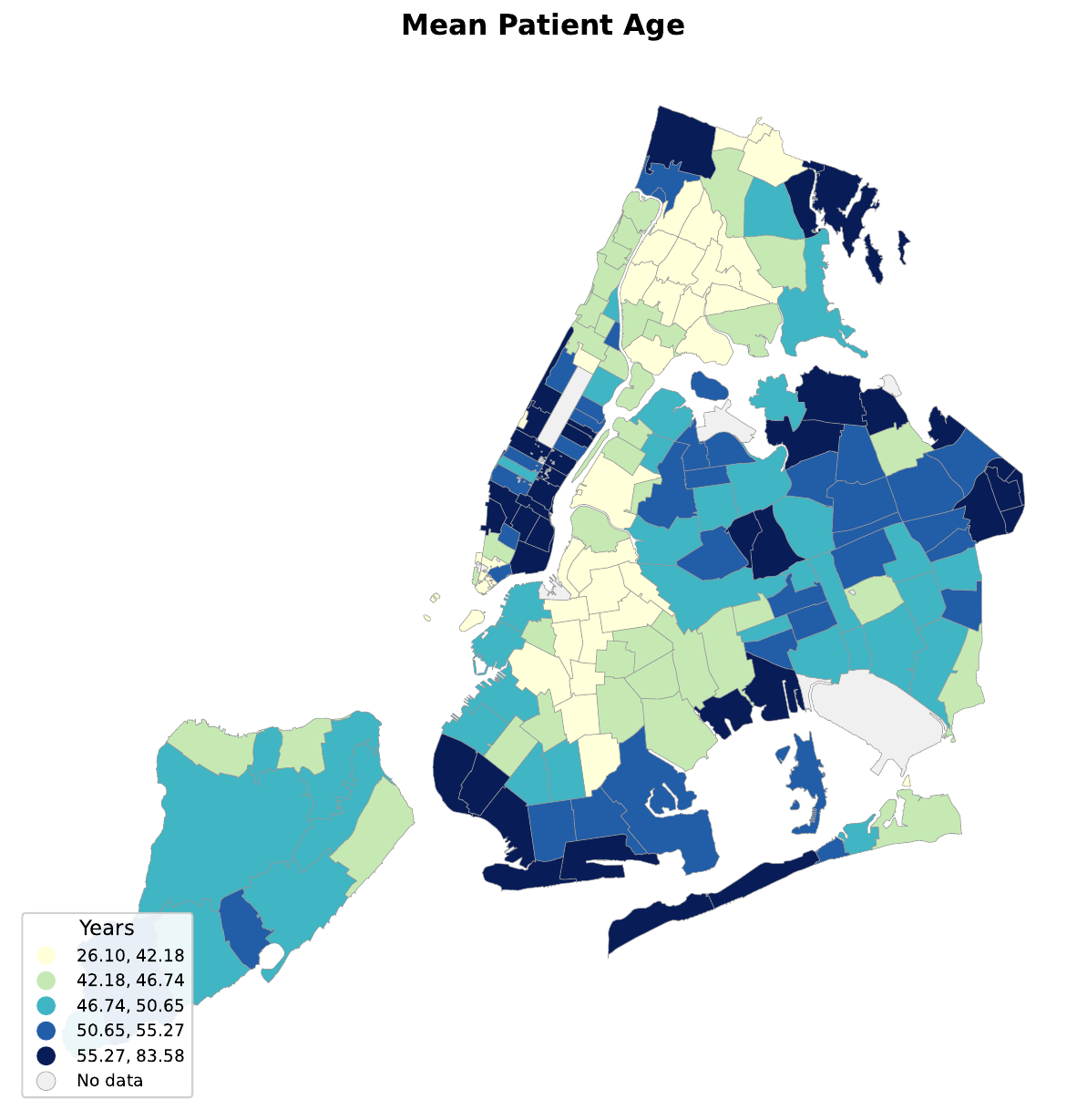}
    \hfill
    \includegraphics[width=0.48\linewidth]{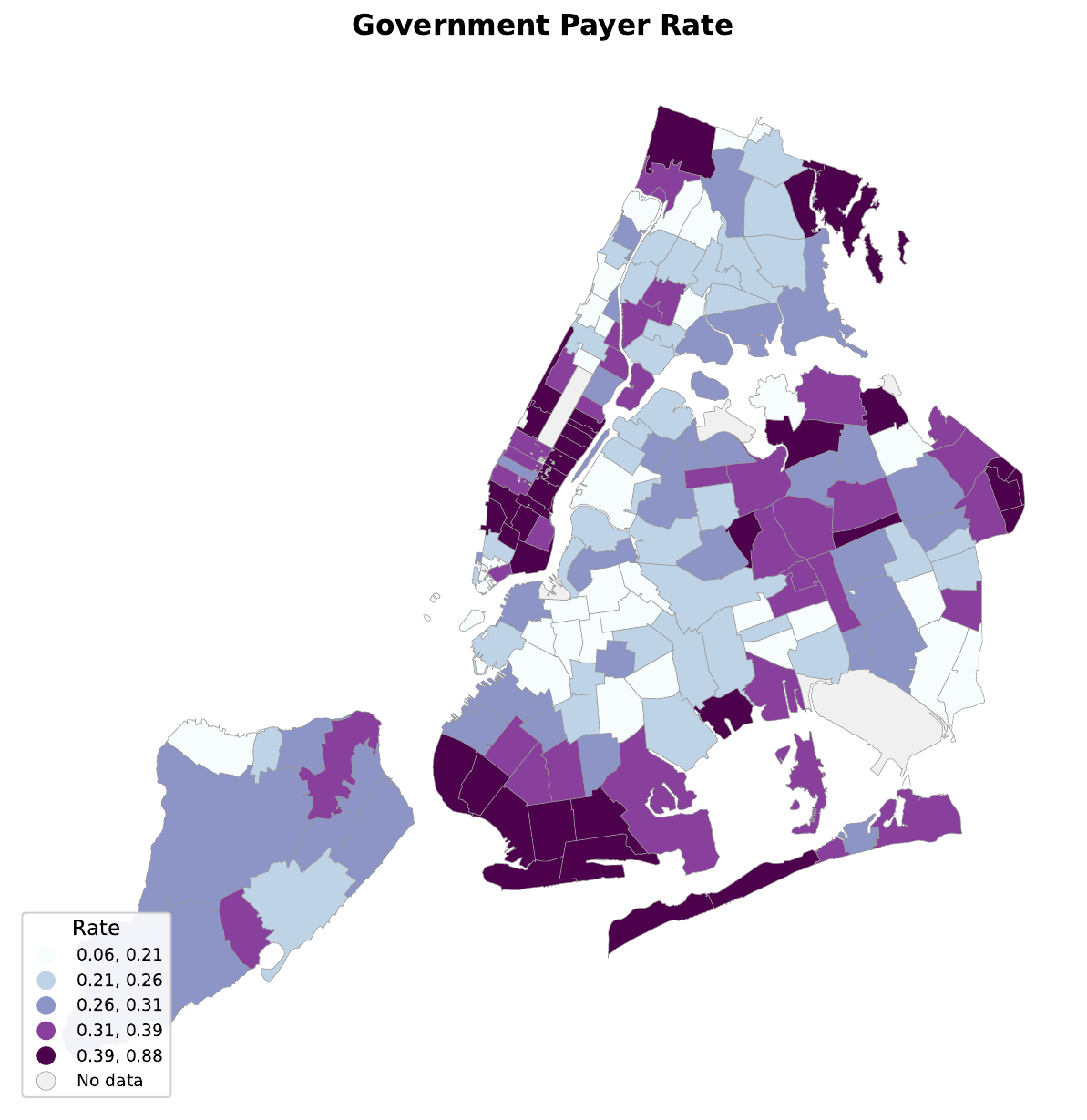}
    \caption{NYC zip code heatmaps of mean patient age (left) and government payer rate (right).}
    \label{fig:nyc_maps_demo2}
\end{figure*}

\begin{figure*}[htp]
    \centering
    \includegraphics[width=0.48\linewidth]{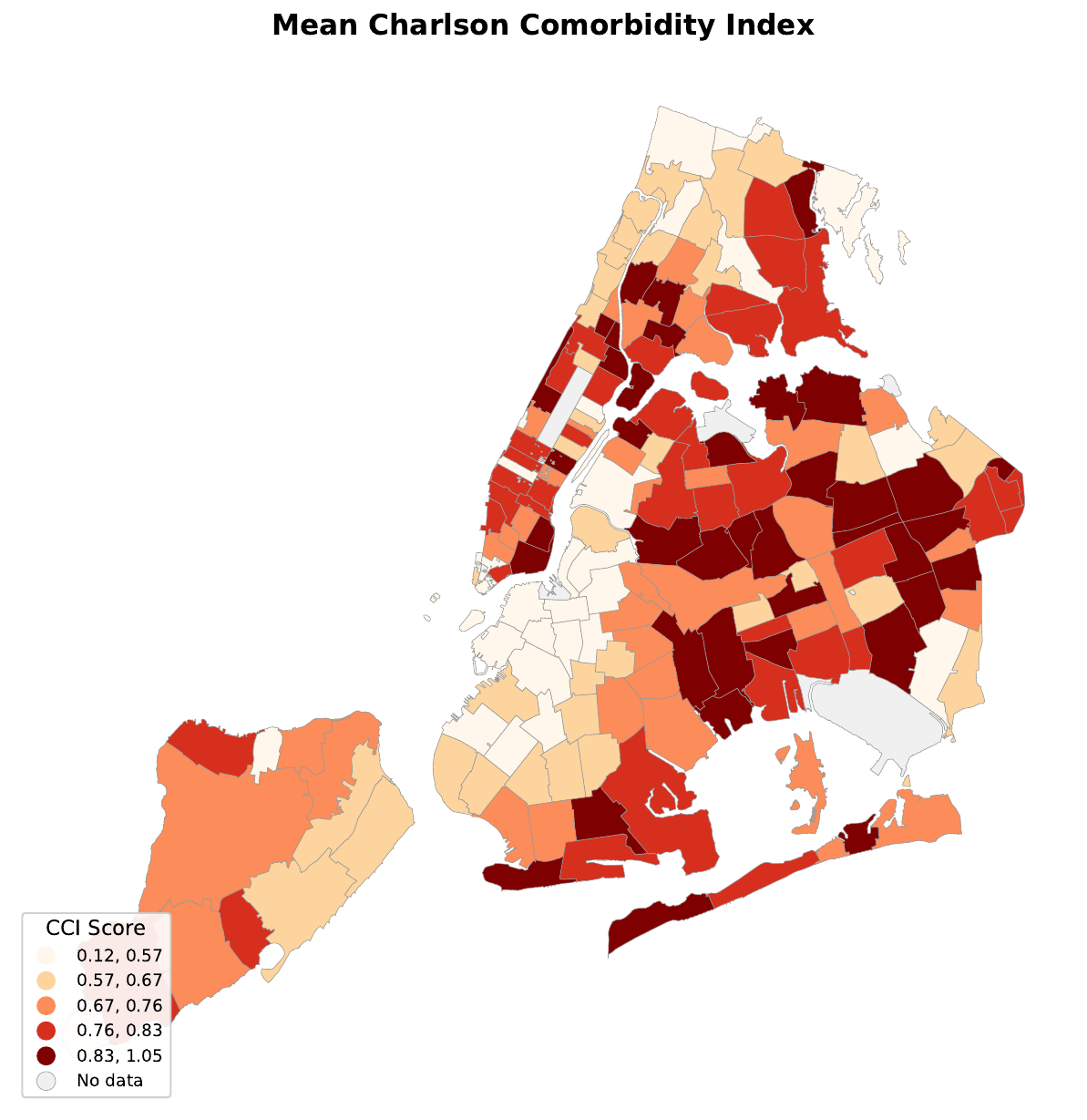}
    \caption{NYC zip code heatmap of mean Charlson Comorbidity Index.}
    \label{fig:nyc_maps_comorbidity}
\end{figure*}

Figures~\ref{fig:nyc_maps_outcomes}--\ref{fig:nyc_maps_comorbidity} show the geographic distribution of health outcomes and demographic attributes across NYC zip codes. These maps illustrate that both clinical outcomes (mortality, length of stay, comorbidity) and demographic attributes (sex, income, age, insurance type) vary substantially by neighborhood, reinforcing the structural link between geography and patient identity exploited by the backdoor paths in \autoref{fig:causal}.

Maps were generated using \texttt{geopandas} and \texttt{matplotlib} by joining per-zip-code aggregate statistics with an NYC zip code shapefile. Aggregates were protected with differential privacy via the Laplace mechanism ($\epsilon = 1.0$, split evenly across columns) and small-group suppression (zip codes with fewer than 10 records were excluded). A quantile classification scheme with five bins was used for visualization.

\end{document}